\documentclass[letterpaper,pra,twocolumn,showpacs,superscriptaddress]{revtex4}
\usepackage{graphicx}
\usepackage{amsmath}
\usepackage{amssymb}
\newtheorem{proposition}{Proposition}
\newtheorem{definition}{Definition}

\begin{document}

\title{Quantum Noise in Multipixel Image Processing}
\author{N. Treps, V. Delaubert, A. Ma\^\i tre, J.M. Courty and C. Fabre}
\affiliation{Laboratoire Kastler Brossel, UPMC, Case 74, 4 Place
Jussieu, 75252 Paris cedex 05, France}

\date{\today}

\begin{abstract}
We consider the general problem of the quantum noise in a
multipixel measurement of an optical image. We first give a
precise criterium in order to characterize intrinsic single mode
and multimode light. Then, using a transverse mode decomposition,
for each type of possible linear combination of the pixels'
outputs we give the exact expression of the detection mode, i.e.
the mode carrying the noise. We give also the only way to reduce
the noise in one or several simultaneous measurements.
\end{abstract}

\pacs{42.50.Dv; 42.30.-d; 42.50.Lc}

\maketitle

\section*{Introduction}

Multipixel photodetectors such as diode arrays or CCD sensors are
now frequently used to record images. These sensors provide
signals in which the useful information is mixed with random
noise. A contribution to this noise originates from the quantum
nature of light : the arrival of individual photons is a random
process. Contrarily to technical noise, due to imperfections in
the source, the optical system or the detector, this quantum noise
can not be reduced by eliminating the defects in the measurement
process. The purpose of this paper is to determine the precise
origin of this noise and to analyze whether and how it can be
reduced. With the analysis of the spatial distribution of this
noise, we will single out the precise transverse modes whose
fluctuations are at the origin of this quantum noise, and
determine the parameters that have to be changed in order to
reduce this noise.

As images are complex objects which carry a great deal of
information, there are actually many ways to extract information
from them, depending on the image user
needs\cite{Bertero,Jain,Katsaggelos}. We will focus our attention  to the extraction from the image of one or
several continuous parameters, the variation of which modifies the
light distribution in the image plane and not its total intensity.
In such a case, the quantity of a priori information on the image
is very important, as one assumes that the variation of the image
under observation is only due to the variation of a searched
parameter $M$. A second use to which our calculations can apply is the determination of predefined patterns in the image, such as
given shapes, surfaces, borders, textures... It is a very
difficult problem per se, and the incidence of quantum noise on
it, to the best of our knowledge, has not been precisely studied
so far. A contrario, we do not consider the
search of the smallest possible details, where resolution is at
stake. In this problem, there is very little to none a priori information and the problem of quantum limits to
resolution has been already considered in other
publications \cite{resolution,kolobov}.

In most cases, the light used to carry the image comes from
"classical sources", such as lamps or usual lasers, in which the
photons are randomly distributed in the image plane. This gives
rise to a spatial shot noise which will yield a "standard quantum
limit" in the measurement of a very small variation of $M$. It is
now well-known that "non-classical light", such as squeezed light
or sub-Poissonian light, is likely to reduce quantum fluctuations
on a given measurement \cite{nonclassical}. The aim of the last
part of the present paper is to identify the best "non-classical
light" enabling us to reduce the quantum noise in the measurement
of the quantity $M$ performed in the image. It has been already
shown \cite{fouet} that non-classical light in a single transverse
mode, though very effective to reduce the noise for a measurement
performed on the total beam, is of little use for a measurement
performed on an image. One therefore needs "multi-transverse-mode
non-classical light" for our purpose. This is the reason why we
devote the first section of this paper to a precise analysis of
such a concept, before considering in the second section the
problem of information extraction : we identify the exact noise
source in the measurement of $M$, and show how to choose the best
configuration which allows us to measure a variation of $M$ with a
sensitivity beyond the standard quantum limit.

\section{"Intrinsic" multimode light}
We consider the propagation of light in the vacuum along the $z$
direction, and call $\vec{r}$ the transverse coordinate.  We
assume that the light frequency is $\omega_{0}$ with a linewidth
$\delta\omega$ much smaller than  $\omega_{0}$, and that it has a
well defined polarization. One knows that it is possible to find
several basis of transverse modes $\{u_i(\vec{r},z)\}$ such as
\begin{itemize}
    \item  each mode verifies the propagation equation of the field in
    vacuum projected onto the polarization axis~:
    \begin{equation}
        \triangle(u_{i}e^{ikz})+\frac{\omega_0^{2}}{c^{2}}u_{i}=0
    \end{equation}

    \item  it is an orthonormal basis :
    \begin{equation}
          \label{eq:base}
          \int u_i^*(z,\vec{r})u_j(z,\vec{r})d^2r=\delta_{ij}
    \end{equation}

    \item it verifies a completeness relation :
    \begin{equation}
        \label{complet}
        \sum_i u_i^*(z,\vec{r})u_i(z,\vec{r}\
        ')=\delta(\vec{r}-\vec{r}\ ')
    \end{equation}
\end{itemize}
For instance, the usual Laguerre-Gauss TEM$_{pq}$ basis verifies
these conditions. Considering a light beam, the electric field is
written as the sum of the positive and negative frequencies
components :
\begin{equation}
    E(\vec r,z,t)=E^{(+)}(\vec r,z)e^{-i(\omega_0t-kz)}+c.c.
\end{equation}
It is possible to expand the electric field positive frequency envelope in the transverse modes basis as
\begin{equation}\label{ChampplusNorm}
  E^{(+)}(\vec{r},z)=\sum_{i}{\cal E}_{i}u_{i}(\vec r,z).
\end{equation}

\subsection{Single mode or multimode light : classical approach}\label{Classical light}

For a TEM$_{pq}$ basis field expansion, when more than one ${\cal
E}_{i}$ is non zero, it seems at first sight natural to say that
this field is multimode. However, if the ${\cal E}_{i}$
coefficients are fixed (i.e. we consider a {\it coherent
superposition of modes} and not a statistical one), one can always
define a new transverse mode :
\begin{equation}
    \label{v0}
    v_{0}=\frac{1}{\sqrt{\sum_{i}|{\cal E}_i|^{2}}}\sum_{i}{\cal E}_{i}u_{i}
\end{equation}
and construct a basis $\{v_{i}\}$ in which $v_{0}$ is the first
element. In this basis, the field is proportional to $v_{0}$ which
means it is single mode. We can conclude that for a coherent
superposition of mode, there is no intrinsic definition of a
multimode beam (i.e. a definition independent of the choice of the
basis). We will restrict our analysis to spatial variables, but it
can be applied to any physical dimension. For instance, in the
time domain, a mode locked laser is single mode, as it is a
coherent superposition of many temporal modes. If the temporal
modes are incoherent with each other then the system is
unambiguously multimode. More precisely, if the field is a
stochastic superposition of modes, the $v_{0}$ mode cannot be
defined and the multimode character has a clear meaning. We will
exclude this case in the following.

\subsection{Single Mode light : quantum approach}
In order to give the quantum description of the transverse plane
of a light beam, it is very common to quantize the field starting
from a transverse mode basis as the one we just defined in the
previous section. In order to obtain standard formulas, we
consider that all measurements are performed in an exposure time
$T$ and associate to each vector of the mode basis a set of
creation and annihilation operators $\hat a_{i}^{\dag}$ and $\hat
a_{i}$ such as the field ${\cal E}_i$ of the previous section is
replaced by the operator
$i\sqrt{\frac{\hbar\omega_0}{2\epsilon_0cT}}\hat a_i$. With these
notations we obtain the standard commutation relations $[\hat
a_i,\hat a^\dagger_j]=\delta_{ij}$, and the positive field
envelope operator can be written as \cite{ClaudeHouches}
\begin{equation}
    \hat E^+(\vec
    r,z)=\sqrt{\frac{\hbar\omega_0}{2\epsilon_0cT}}\hat A^+(\vec
    r,z)
\end{equation}
with
\begin{equation} \label{Qbase}
    \hat A^{(+)}(\vec r,z)= \sum_{i}\hat a_{i}(z)u_{i}(\vec r,z)
\end{equation}
so that $\hat A^{(+)\dagger}\hat A^{(+)}$ is a photon number per
unit surface.

In order to give a proper definition of a single mode case, let's
write the most general state of the field in the Fock state basis
$|n_{1},\ldots,n_{i},\ldots\rangle$, where $n_{i}$ stands for the
number of photons in the mode $i$ :
\begin{equation}
    |\psi\rangle=\sum_{n_{1},\ldots,n_{i},\ldots}C_{n_{1},\ldots,n_{i},\ldots}
    |n_{1},\ldots,n_{i},\ldots\rangle
\end{equation}
and the mean value of the electric field is given by :
\begin{eqnarray}
    \langle\psi|\hat A|\psi\rangle & = & \sum_{i}
    \left(\sum_{n_{1},\ldots,n_{i}>1,\ldots}\right. \nonumber \\
    & & \left. C^{*}_{n_{1},\ldots,n_{i}-1,\ldots}
    C_{n_{1},\ldots,n_{i},\ldots}\right)
    \sqrt{n_{i}}u_{i}(\vec r)
\end{eqnarray}
Following the definition for the classical beams, we can give a
definition of a single mode beam :
\begin{definition}\label{defsingle}
A state is single mode if it exists a mode basis
$\{v_{0},v_{1},\dots \}$ in which it can be written :
\begin{equation*}
    |\psi\rangle=|\phi\rangle\otimes|0,\ldots,0,\ldots\rangle
\end{equation*}
where $|\phi\rangle$ is the state of the field in the first
transverse mode.
\end{definition}
The question is now whether, in contrast with the classical
states, it exists quantum states that cannot be written as
(\ref{defsingle}). To answer this question, we will demonstrate
the following proposition :

\begin{proposition}\label{monomode}
    A quantum state of the field is single mode if and only if the action on it of all
    the annihilation operators of a given basis give
    collinear vectors.
\end{proposition}
On can note that if this property stands for a given basis, it then stands for the
action of any annihilation operator.

Let's assume first that our field $|\psi\rangle$ is single mode
with respect to the basis $\{u_{i},\hat a_{i}\}$, then
\begin{equation}
    \hat a_{0}|\psi\rangle=|\psi_{0}\rangle \quad \textrm{and} \quad
    \hat a_{i}|\psi\rangle=0 \quad \forall i\neq 0
\end{equation}
Consider now any linear combination of the operators
\begin{equation}
    \hat b=\sum_{i}c_{i}\hat a_{i}
\end{equation}
where $\sum_{i}|c_{i}|^2=1$ which ensures that $[\hat b,\hat b^{\dagger}]=1$. The action of this operator on the field is given by :
\begin{equation}
    \hat b|\psi\rangle=\sum_{i}c_{i}\hat
    a_{i}|\psi\rangle=c_{0}|\psi_{0}\rangle
\end{equation}
This demonstrates the first implication of our proposition~: all the
actions of annihilation operators on the field are proportional.

To prove the other implication, consider now a field
$|\psi\rangle$ on which the action of any annihilation operator
$\hat a_{i}$ is proportional to $|\psi_{0}\rangle$. This is in
particular true for the basis $\{u_{i},\hat a_{i}\}$ :
\begin{equation}
    \hat a_{i}|\psi\rangle=\alpha_{i}|\psi_{0}\rangle
\end{equation}
If we assume that $\sum_{i}|\alpha_{i}|^{2}=1$ (which is always
possible by changing the normalization of $|\psi_{0}\rangle$), we
can define a new basis $\{v_{i}(\vec r,z),\hat b_{i}\}$ such as
\begin{equation}
    \label{bonmode}
    \hat b_{0}=\sum_{i}\alpha_{i}^{*}\hat a_{i}, \quad
    v_{0}=\sum_{i}\alpha_{i}^{*}u_{i}
\end{equation}
and complete the basis by defining
a unitary matrix $[c_{ij}]$ such as
\begin{equation}
    \hat b_{i}=\sum_{j}c_{ij}\hat a_{j} \quad \textrm{with }
    c_{0j}=\alpha_{j}^{*} \quad \textrm{and }
    \sum_{j}c_{ij}c_{kj}^{*}=\delta_{ik}.
\end{equation}
It is then straightforward to show that
\begin{equation}
    \hat b_{i}|\psi\rangle=\delta_{0i}|\psi_{0}\rangle
\end{equation}
which concludes the demonstration.\\

In addition to the proposition, equation (\ref{bonmode}) gives the
expression of the mode in which "lies" the mean field, knowing the action of a particular basis. We can also note that to show
that a field is single mode, it is sufficient to show that all its
projections on the annihilation operators of one particular basis
are proportional.

To illustrate the proposition, if one considers the superposition
of coherent states
\begin{equation}
    |\psi\rangle=|\alpha_{1}\rangle\otimes\ldots\otimes|\alpha_{i}\rangle\otimes\ldots
\end{equation}
it is straightforward to show that the actions of all the annihilation operators on this state are
proportional to the state itself ; we have a single mode beam.
The basis in which it is single mode is the same as the one for
the classical case, setting $v_{0}$ as in equation (\ref{v0}).

Using this proposition, we can also look for the different states
that fulfill our definition of single mode quantum beam. As a
state that cannot be written as follows in any mode basis :
\begin{equation}
    |\psi\rangle=|\phi_{1}\rangle\otimes\ldots\otimes|\phi_{i}\rangle\otimes
    \ldots
\end{equation}
is obviously not a single mode beam, we will consider now such a
factorized state of the field, on which the action of the
annihilation operators gives :
\begin{equation}
    \hat a_{i}|\psi\rangle=|\phi_{1}\rangle\otimes\ldots\otimes(\hat
    a_{i}|\phi_{i}\rangle)\otimes\ldots
\end{equation}
Consequently, there are only two possibilities to have all these
states proportional :
\begin{itemize}
    \item  either only one of the action is different from zero,
    which means we are already in the basis in which the state is
    single mode.
    \item  or all the states are coherent states.
\end{itemize}
We have described here all the possible single mode states, which
agree with the intuitive description one can have. For instance,
if one considers the superposition of several transverse modes, if
at least one of them is a non-coherent state, one gets a quantum
multimode state.

\subsection{Multimode light : quantum approach}\label{Multimode quantum light}
A beam of light is said multimode, from a quantum point of view,
when it is not single mode according to definition \ref{defsingle}. We can
characterize such a beam by its degree $n$ (this degree equals one
for a single mode beam):

\begin{definition}
  For a beam $|\psi\rangle$, the minimum number of modes necessary to
  describe it (or the minimum number of non-vacuum modes in its modal
  decomposition), reached by choosing the appropriate basis, is called
  the degree $n$ of a multimode beam. Any corresponding basis is
  called a minimum basis for the field $|\psi\rangle$.
\end{definition}

The degree of a multimode beam can also be related to the
generalization of proposition (\ref{monomode}) to a $n$-mode beam.
Using the same technique, one can show that a quantum field is a
$n$-mode beam if and only if the action on it of all the
annihilation operators belongs to the same $n$ dimensioned
sub-space.

Whereas the previous paragraph gives a good definition of the
degree of a multimode beam, it is not very convenient as one has
no information on the basis in which the beam is exactly described
by $n$ modes. We can however define a particular basis, useful for
calculations :
\begin{proposition} \label{eigenbasis}
    For a beam $|\psi\rangle$ of degree $n$, it is always possible to
    find a basis $\{u_{i},\hat a_{i}\}$ such as
    \begin{itemize}
        \item The mean value of the electric field is non zero only in
          the first mode.
        \item It is a minimum basis for the field $|\psi\rangle$.
    \end{itemize}
    we will call that basis an eigenbasis.
\end{proposition}

In order to demonstrate this proposition, let's consider a minimum
basis $\{u_{i},\hat a_{i}\}$ for the field $|\psi\rangle$. This
basis is supposed to be ordered such as the $n$ first modes are
the relevant ones. We can then define a new basis $\{v_{i},\hat
b_{i}\}$ such as :
\begin{eqnarray} \label{eigenbase}
    v_{0} & = & \frac{1}{\sqrt{\sum_{i=0}^{n-1}\langle\hat
    a_{i}\rangle^{2}}}\sum_{i=0}^{n-1}\langle\hat a_{i}\rangle u_{i}
    \nonumber \\
    v_{i,\ 0<i<n} & = & \sum_{j=0}^{n-1}c_{ij}u_{j} \nonumber \\
    v_{i,\ i\geq n} & = & u_{i}
\end{eqnarray}
where the coefficient $\{ c_{ij}\}$ are chosen in order to get an
orthonormal basis. Definitions similar to the one of equ.
(\ref{eigenbase}) apply for the annihilation operators. The first
vector of this basis is the same as the one defined for a
classical beam in equation (\ref{v0}). In that basis, the mean
field is single mode in a classical sense. However the energy
lying in all the other modes is not necessarily zero, only the
electric field mean value is zero for these modes, and as the
modes for $i\geq n$ were not changed, this new basis is still a
minimum one for the field $|\psi\rangle$. This demonstrates the
proposition. The demonstration illustrates the construction of a
basis as defined in proposition (\ref{eigenbasis}) from a minimum
basis, even though thanks to the $\{ c_{ij}\}$ coefficients an
infinite number of basis are possible.

The existence of this basis is also a confirmation of the
intuitive idea of the difference between single mode and multimode
quantum light. Indeed, for a single mode beam, the spatial
variation of the noise is the same as the one of the mean field.
For a multimode beam, the previous description shows that some of
the modes orthogonal to the mean field are sources of noise but do
not contribute to the mean field. This implies that the variation
of the noise is independent of the one of the mean field. This
property can be used to experimentally characterize the multimode
character of light. One of the difficulties of such experiments is
the knowledge of the mode structure of the field, as it is not
possible to test all the transverse modes. We have shown that
\cite{Marcelo} the quantum multimode character of light can be
demonstrated using an iris whose aperture size is continuously
varied. We applied it to the case of quantum correlated light
coming out of an optical parametric oscillator.

We have defined the theoretical basis required to develop a study
on optical image measurements. The following section on
information extraction will indeed strongly rely on the
propositions and definitions of the first part.

\section{Difference Measurements}\label{dm}

\subsection{Description}
A widely used technique in optics, and more generally in physics,
to improve the signal to noise ratio in a measurement is to
perform a \textit{difference measurement} : it consists in
producing two identical signals from the light source used in the
experiment. When one monitors the difference between these two
signals, one gets of course a zero mean signal, but one also
cancels all the common mode noises, for example the one arising
from the classical intensity fluctuations of the source. The
remaining noise arises from the noise sources affecting
differently the two channels.

One simple way to produce the two identical beams is to use a
$50\%$ beamsplitter. In this case, the vacuum noise coming from
the unused side of the splitter is such a not-common mode noise
and remains on the difference measurement : whatever the actual
excess noise of the beam impinging on the beam splitter, the
remaining noise corresponds to the shot noise of this beam.

This simple technique of noise cancellation is used for example to
measure very small absorptions \cite{Schwob} by inserting the
absorbing medium in one of the arms of the difference set-up, or
very small frequency shifts, by inserting a Fabry-Perot cavity in
one of the arms. It is also extensively used in multipixel
measurements, either with split-detectors or quadrant detectors,
to measure sub-micrometer displacements, for example of nanoscale
fluorophores in biological samples \cite{biologie} and  in Atomic
Force Microscopy \cite{Senden}, and ultra-small absorptions by the
mirage effect \cite{Boccara}.

The problem of the determination of the origin of quantum noise on
a split detector and of its reduction has been already
investigated theoretically \cite{fouet} and experimentally
\cite{Prl, Science, longspatial}. We will here extend these
considerations to more general configurations.


More formally, we consider the measurement by a detector consisting of a
set of pixels, each one occupying a transverse area $D_{i}$. The pixels cover the whole transverse plane, with no overlap
between them. Each photodetector delivers a power given by~:
\begin{equation}
    \hat I(D_{i})=\int_{D_{i}}2\epsilon_{0}c\hat E^{\dagger}(\vec r)\hat
    E(\vec r)d^{2}r
\end{equation}
This can also be written in photon number measured during the
exposure time $T$ of the detector :
\begin{equation}
    \hat N(D_{i})=\int_{D_{i}}\hat A^{\dagger}(\vec r)\hat
    A(\vec r)d^{2}r
\end{equation}
In this section, the measurement $M$ is defined by :
\begin{equation} \label{equalgain}
    \hat M(\{\sigma_{i}\})=\sum_{i}\sigma_{i}\hat I(D_{i}) \quad
    \textrm{such as } \sigma_{i}=\pm 1
\end{equation}
or again in terms of number of photons per second :
\begin{equation} \label{equalgain2}
    \hat N(\{\sigma_{i}\})=\sum_{i}\sigma_{i}\hat N(D_{i})
\end{equation}
 where
$\sigma_{i}=\pm 1$ corresponds to the electronic gain of detector {\it i}.

Considering a light beam in state $|\psi\rangle$, the measurement
is a difference measurement for that beam if its mean value is
zero, i.e. if
\begin{equation} \label{differential}
    \langle\hat N(\{\sigma_{i}\})\rangle=0
\end{equation}

\subsection{One difference measurement}\label{differential measurement}

If one consider one difference measurement performed with a
coherent state, which has spatially uncorrelated quantum
fluctuations, the noise arising from the measurement will not
depend on the choice of $\{\sigma_{i}\}$ if $\sigma_{i}=\pm 1$,
and will be equal to the square root of the total number of
photons. This is what is called the standard quantum
noise. In the general case, in order to compute the noise, an
analysis equivalent to the one performed in the case of a small
displacement measurement, as done in reference \cite{fouet}, is
necessary. We recall it here and extend it to the general case of
transverse modes of any shape, in order to show the following
proposition :
\begin{proposition}
The noise on a difference measurement performed on a beam
    $|\psi\rangle$ originates from a single mode, orthogonal to the mean field : the "flipped mode".
    In order to reduce the noise in that measurement, it is
    necessary and sufficient to inject a squeezed state in this flipped mode.
\end{proposition}

In order to perform the general noise calculation, let us define
the two "detectors" :
\begin{eqnarray}
    D_{+} & = & \bigcup_{i,\sigma_i=+1}D_{i} \nonumber \\
    D_{-} & = & \bigcup_{i,\sigma_i=-1}D_{i}
\end{eqnarray}
which gives
\begin{eqnarray}
    \hat N_{-} & = & \hat N(D_{+})- \hat N(D_{-})\nonumber \\
    & = &
    \sum_{i,j}\hat{a}^{\dagger}_{i}
    \hat{a}_{j}\bigg[\int_{D_{+}}u_{i}^{*}(\vec r)u_{j}(\vec r)d^{2}r\nonumber\\
    && \qquad - \int_{D_{-}}u_{i}^{*}(\vec r)u_{j}(\vec r)d^{2}r\bigg]
\end{eqnarray}
Considering small quantum fluctuations for which $\delta\hat
a_{i}=\hat a_{i}-\langle \hat a_{i}\rangle$,  the fluctuations of
$\hat N_{-}$ are
\begin{eqnarray}
    \delta \hat N_{-} & = & \hat N_{-}-\langle\hat N_{-}\rangle
    \nonumber\\
    & = & \sum_{i}\big[\delta\hat
a_{i}^{\dagger}C^{i}_{-}+\delta\hat a_{i}C^{i*}_{-}\big]
\end{eqnarray}
with $C_{-}^{i}$ defined as
\begin{eqnarray}
    C_{-}^{i}&=&\sum_{j}\langle\hat a_{j}\rangle\Big[\int_{D_{+}}u_{i}^{*}(\vec r)u_{j}(\vec r)d^{2}r-
    \int_{D_{-}}u_{i}^{*}(\vec r)u_{j}(\vec r)d^{2}r\Big]\nonumber\\
    &=& \int_{D_{+}}u_{i}^{*}(\vec r)A_{\psi}(\vec
    r)d^{2}r-\int_{D_{-}}u_{i}^{*}(\vec r)A_{\psi}(\vec r)d^{2}r\nonumber
\end{eqnarray}
and where $A_{\psi} (\vec r)$ is the mean value of the electric
field $\langle\psi|\hat
 A(\vec r)|\psi\rangle$. The $C_{-}^{i}$ coefficients are the
 partial overlap integrals between the modes $u_{i}$ and the mean field.

We can now compute the noise related to this measurement :
\begin{eqnarray}\label{noise}
    \langle\delta\hat N_{-}^{2}\rangle=\sum_{i}|C_{-}^{i}|^{2}
     & + & \bigg[\sum_{i,j}\langle\delta\hat a_{i}^{\dagger}\delta\hat
    a_{j}^{\dagger}\rangle C_{-}^{i}C_{-}^{j} \nonumber \\
    & &  + \langle\delta\hat a_{i}^{\dagger}\delta\hat
    a_{j}\rangle C_{-}^{i}C_{-}^{j*}+c.c.\bigg]
\end{eqnarray}
Using the completeness relation, the first term of the last
equation can be shown to be equal to the total number of incident
photons per second $N_{0}$. This shows that the noise
related to this measurement arises a priori from all the modes.

We will now demonstrate that the noise comes in fact from a single
mode when we write $\langle\delta\hat N_{-}^{2}\rangle$ in the
appropriate basis.

We call $v_{0}$ the mode of the mean field as defined in the
previous part :
\begin{equation}
    v_0(\vec r)=\frac{1}{\sqrt{N_0}}A_{\psi}(\vec r).
\end{equation}
If $v_0$ is the first mode of a basis, the mean value of the
electric field in all the other modes will be zero, as shown in
the previous section. We define now the mode $v_{1}$, which we
will refer to as the "flipped mode" of $v_{0}$, such as :
\begin{eqnarray} \label{flipped}
  v_{1}(\vec r) & = & v_{0}(\vec r) \quad \textrm{if } r\in
  D_{+} \nonumber \\
  v_{1}(\vec r) & = & -v_{0}(\vec r) \quad \textrm{if } r\in D_{-}
\end{eqnarray}
As we have assumed that the mean value of the measurement is zero,
$v_1$ is orthogonal to $v_0$, which means that we can find a basis $\{v_i,\hat b_i\}$ where $v_0$ and $v_1$ are the two first modes. In that basis, the
overlap integrals become :
\begin{eqnarray}
    C_{-}^{i} &=& \sqrt{N_{0}}\Big[\int_{D_{+}}v_{i}^{*}(\vec r)v_{0}(\vec r)d^{2}r-\int_{D_{-}}v_{i}^{*}(\vec r)v_{0}(\vec r)d^{2}r\Big]\nonumber \\
    &=& \sqrt{N_{0}}\int_{D}v_{i}^{*}(\vec r)v_{1}(\vec r)d^{2}r
    \nonumber \\
    &=& \sqrt{N_{0}}\delta_{i,1}
\end{eqnarray}
These integrals are different from zero only for the flipped mode.
The noise of equation (\ref{noise}) becomes
\begin{equation}\label{result}
   \langle\delta\hat N_{-}^{2}\rangle = N_{0}\langle(\delta\hat
   b_{1}^{\dagger}+\delta\hat b_{1})^{2}\rangle
\end{equation}
which shows that the noise only arises from the quadrature of the
flipped mode of $v_{0}$ in phase with the mean field mode. For
this reason, we call this mode the eigenmode of the measurement. An other standard notation is
\begin{eqnarray}
   \langle\delta\hat N_{-}^{2}\rangle = N_{0}\langle\delta X_
   {1}^{+2}\rangle
\end{eqnarray}
where $X_{1}^{+}=\hat b_{1}+\hat b_{1}^{\dagger}$ is the
quadrature of the flipped mode, and $N_{0}$ represents the shot
noise. Consequently, having a squeezed state in that mode is necessary and sufficient to reduce
the noise related to the measurement.
%

This calculation shows that, for a difference measurement, the
noise in the measurement is exactly the one of the flipped mode.
Changing the noise properties of the flipped mode is then the only
way to change the noise in the measurement. We have a necessary
and sufficient condition to improve the measurement compared to
the standard quantum limit.

This demonstration imposes the noise properties of only one
quadrature of the flipped mode, but there is no condition on the
other quadrature, and all the other modes can be in any state.
Then, there is not only one practical solution.

\subsection{Multiple difference measurement}\label{multidifferential}
We have demonstrated which mode one needs to squeeze in order to
perform one difference measurement on a beam. We can now expand
this analysis in the case of several difference measurements. Let
us consider $n$ difference measurements of the type of equation
(\ref{differential}). We will assume that these measurements are
independent, which means that none of them is a linear combination
of the other ones. One can show that the corresponding flipped
modes are then also linearly independent. We have shown that in
order to improve simultaneously the sensitivity of all these
measurements it is necessary, and sufficient, to squeeze all these
flipped modes. Practically these modes are in general not
orthogonal, but one can find an orthogonal basis of the subspace
generated by these modes. Injecting squeezed vacuum states in each
of these modes will result in squeezed states in each of the
flipped modes.

Regarding the degree of the beam necessary to improve
simultaneously all the measurements, it is clear that in order to
perfectly squeeze all the flipped modes, a beam of degree $n+1$ is
necessary (and sufficient). We can summarize all the
considerations of section \ref{dm} into a proposition~:

\begin{proposition}
In order to reduce the noise simultaneously in n independent
    difference measurements it is necessary and sufficient
    to use a beam of degree at least n+1 that can be described in a
    transverse mode basis $\{\hat a_{i},u_{i}\}$ such as : $u_{0}$
    is proportional to the electric field profile of
        the beam; $\{u_{i}\}_{0<i\leq n}$ is the basis of the space-vector
        generated by the flipped modes of the measurements and
        all these modes are perfectly squeezed.
\end{proposition}

\section{Linear measurement}
Difference measurements are obviously not the only ones performed
in image processing \cite{Bertero,Jain,Katsaggelos}. The
extraction of the pertinent information arises generally from the
numerical computation of a function $F\left(I(D_1),I(D_2),...,
I(D_n)\right)$ from the intensities $I(D_i)$ ($i=1,...,n$)
measured on each pixel. To simplify the following discussion, we
will restrict ourselves to the case when this function is
\textit{linear} with respect to the intensities $I(D_i)$, as it is
a case often encountered in real situations, for example when one
wants to determine the spatial Fourier components of the image, or
when the variations of the parameter to measure are small enough
so that the function $F$ can be linearized.

In the formalism of equations (\ref{equalgain})
and(\ref{equalgain2}), using a linear function correspond to let
the gain $\sigma_i$ of the detectors take any real value and not
only $\pm 1$~:
\begin{eqnarray}
    \hat M(\{\sigma_{j}\})& = & \sum_{j}\sigma_{j}\hat I(D_{j}) \nonumber \\
    \hat N_\sigma & = & \sum_{j}\sigma_{j}\hat N(D_{j})
\end{eqnarray}
We emphasize that, contrary to the previous section, the mean
value of the measurement is not necessarily zero. In that case, we
will show the following proposition :
\begin{proposition}
   Considering a field state $|\psi\rangle$ described in an eigenbasis
   $\{\hat b_i,v_i\}$, and considering a linear measurement performed
   with and array of detectors ${D_i}$, each detector having a gain
   $\sigma_i$. The noise on the measurement $\hat N_\sigma=\sum_j \sigma_j\hat N(D_j)$
   arises only from the generalized flipped mode $w$ defined by :
\begin{equation}
\forall \vec r,\ \vec r \in D_{i}\  \Rightarrow \ w_{1}(\vec
r)=\frac{1}{f}\sigma_{i}v_{0}(\vec r)
\end{equation}
where $f$ is a normalisation factor.
\end{proposition}

Here, there is not much sense in defining the positive and
negative gain domains. We can anyway extend the notion of overlap
integral between a basis vector and the mean field :
\begin{equation}\label{Cgeneral}
    C_{\sigma}^{i}=\sum_{j}\sigma_{j}\int_{D_{j}}u_{i}^{*}(\vec r)A_{\psi}(\vec r)d^{2}r
\end{equation}
which leads to a formula equivalent to equation (\ref{noise})
\begin{eqnarray}\label{noisesigma}
    \langle\delta\hat N_\sigma^{2}\rangle=\sum_i|C_\sigma^i|^2
     & + & \bigg[\sum_{i,j}\langle\delta\hat a_{i}^{\dagger}\delta\hat
    a_{j}^{\dagger}\rangle C_\sigma^{i}C_\sigma^{j} \nonumber \\
    & &  + \langle\delta\hat a_{i}^{\dagger}\delta\hat
    a_{j}\rangle C_\sigma^{i}C_\sigma^{j*}+c.c.\bigg]
\end{eqnarray}

Recalling that $A_{\psi}(\vec r)=\sqrt{N_{0}} v_{0}(\vec r)$, we
can also extend the notion of flipped mode, and define a {\it
detection mode} by
\begin{equation}\label{generalflip}
\forall \vec r,\ \vec r \in D_{i}\  \Rightarrow \ w_{1}(\vec
r)=\frac{1}{f}\sigma_{i}v_{0}(\vec r)
\end{equation}
where $f$ ensures the normalization of $w_{1}$ :
\begin{equation}
    f^{2}=\sum_{j}\sigma_{j}^{2}\int_{D_{j}}v_{0}^{*}(\vec r)v_{0}(\vec r)d^{2}r.
\end{equation}
However, as the mean value of the measurement can be different
from zero, the detection mode $w_1$ is not in general orthogonal
to the mean field mode $v_0$. In order to calculate the noise in
the measurement, it is necessary to construct a basis that contain
the detection mode $w_1$. As the mean value of the electric field
in this mode is different from zero, it is not possible to obtain
an eigenbasis with $w_1$, but we can still choose $w_0$ such as
the mean field mode $v_0$ is a linear combinaison of $w_0$ and
$w_1$. Choosing all the other modes $w_{i}$ (with $i\geq 2$) in
order to obtain an orthonormal basis, we obtain a basis such as
the mean field is distributed in the two first modes, the
detection mode is $w_1$ and the mean value of the electric field
in all the other modes is zero. We can then perform a calculation
similar to the one of the previous section that gives~:
\begin{equation}
  C_{\sigma}^{i}=\sqrt{N_{0}}f\int_{D}w_{i}(\vec r)^{*}w_{1}(\vec r)d^{2}r=\sqrt{N_{0}}f \delta_{i,1}
\end{equation}
Once again the detection mode is the only one that is relevant for
the calculation of the noise related to the measurement. Taking
into account that the normalization giving rise to the shot noise
has changed :
\begin{equation}
   \sum_{i}
   |C_{\sigma}^{i}|^{2}=|C_{\sigma}^{1}|^{2}=N_{0}f^{2}
\end{equation}
the noise formula becomes :
\begin{equation} \label{gennoise}
   \langle\delta\hat
   N_\sigma^{2}\rangle=f^{2}N_{0}\langle(\delta\hat
   c_{1}^{\dagger}+\delta\hat c_{1})^{2}\rangle
\end{equation}
where the $\{\hat c_i\}$ are the annihilation operators associated with the transverse mode basis $\{w_i\}$.

The $f^{2}$ factor is a global effect of the gain, and modifies both the
measured signal and shot noise level. In any case, if the flipped
mode is perfectly squeezed, we can still perform a perfect
measurement. However, the experimental configuration is much more
complicated as, in general, the mean value of the electric field
in mode $w_1$ is different from 0, which means that, as is shown in the appendix, generating the good mode is difficult. An appropriate approach would be to
describe the field back into an eigenbasis, and check how to set
the noise of the different modes in that basis. We will see in
appendix \ref{appendix} how this can be done in a simple case. The important result of this part is that whatever the
measurement we perform the noise arises only from one mode.
Changing the noise of this mode allows us to improve the sensitivity
of the measurement. As in the previous section, it is also
possible in that general case to perform several simultaneous
measurements, and to identify the subspace of modes responsible
for the noise.

It is interesting to note that, in the particular case of a
measurement where the gains are adapted to have $\langle\hat
M(\{\sigma_{j}\})\rangle = 0$, the mode $v_{0}$ coincides with
$w_{0}$. Indeed, $v_{0}$ is here orthogonal to $w_{1}$ :
\begin{eqnarray}
    \int_{D}w_{1}^{*}(\vec r)v_{0}(\vec r)d^{2}r &=& \sum_{j}
    \frac{\sigma_{j}}{f}\int_{D_{j}}v_{0}^{*}(\vec r)v_{0}(\vec r)d^{2}r \\ \nonumber
    & \propto & \langle\sum_{j}\sigma_{j}\hat N(D_{j})\rangle \\\nonumber
    &=& 0
\end{eqnarray}
hence the basis is an eigenbasis of the field. Again, that case is
relevant experimentally as it means that one can act on the noise
without perturbing the mean field mode.

\section*{Conclusion}

We have shown in that article how to properly define the degree of
multimode character of a light beam. We have used the basis
decomposition associated with that definition in order to single
out, in a linear transverse measurement, the transverse mode
carrying the noise. We have shown that it was possible to go
beyond the standard quantum noise limit by injecting in that mode
squeezed light, and that this could be done simultaneously for
several independent measurement.

It order to implement the theory developed here to complex experimental configurations we have shown that it was preferable that the various detection modes be orthogonal to the mean field (i.e. they do not contribute to the mean electric field), and it is necessary to mix them without introducing losses. For instance, one can use the proposal we have detailled in \cite{longspatial} and used to mix two non-classical beams in orthogonal transverse modes, and a mean coherent field, in order to improve the sensitivity of the transverse position measurement of a laser beam.

In this paper, we have analyzed in great detail the origin of
quantum noise in a multi-pixel measurement. What remains to be
considered now is the signal, and not only the noise in the
measurement. This will be the natural continuation of our work,
and we will describe in a future publication what is the influence
of the gain configuration to the signal to noise ratio and how to
optimize a given measurement in an optical image.

\section*{Acknowledgments}

Laboratoire Kastler Brossel, of the Ecole Normale Sup\'{e}rieure
and University Pierre et Marie Curie, is associated to CNRS. This
work has been supported by the European Union in the frame of the
QUANTIM network (contract IST 2000-26019).

\appendix
\section*{Appendix : Two zone measurement}\label{appendix}

In this article, we have exhibited the mode structure of the light
in a multipixel measurement, using a basis that contain the
detection mode. However, when the mean value of the measurement is
different from zero, we have shown that this detection mode has a
mean electric field value different from zero. In that
configuration, it is very difficult experimentally to address the
detection mode without modifying the mean field distribution. We
have shown that the only basis pertinent for such a task was an
eigenmode basis, we will show here what is the structure of that
basis for a two zone measurement of non-zero mean value.

Using the notations of the previous sections, we consider two
detectors $D_+$ and $D_-$ whose gains are respectively $+1$ and
$-1$. We recall here the mode structure defined in the main text
of that article. $v_{0}$ is the transverse mode carrying the mean
field of the beam and $w_{1}$ is the detection mode as define in
equation (\ref{generalflip}) (which, in that case is equivalent to
the flipped mode of equation ({\ref{flipped})). $w_{0}$ is the
mode orthogonal to $w_{1}$ in the subspace generated by $v_{0}$
and $w_{1}$. Let us call $i_{+}$ and $i_{-}$ the partial integrals
of $v_{0}$ on each zone,
\begin{equation}
    i_{+}=\int_{D_{+}}v_{0}^{*}(\vec r)v_{0}(\vec r)d^{2}r \quad \textrm{and
    }\quad i_{-}=\int_{D_{-}}v_{0}^{*}(\vec r)v_{0}(\vec r)d^{2}r \nonumber
\end{equation}
a simple calculation gives
\begin{eqnarray}
    w_{0}(\vec r) & = & \sqrt{\frac{i_{-}}{i_{+}}}v_{0}(\vec r) \quad \textrm{if } r\in D_{+} \nonumber \\
    w_{0}(\vec r) & = & \sqrt{\frac{i_{+}}{i_{-}}}v_{0}(\vec r) \quad \textrm{if } r\in D_{-}
\end{eqnarray}
\begin{figure}
    \centerline{\includegraphics[width=8.4cm]{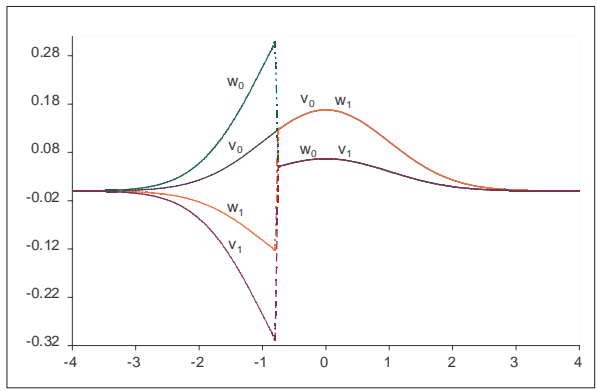}}
    \caption{Electric field profile of the constituent modes used
    to form the non-classical multimode beam.}
    \label{fig}
\end{figure}
The first mode of an eigenbasis for the field is $v_0$. The second
one, $v_1$ is defined as the mode orthogonal to $v_0$ in the
subspace generated by $w_0$ and $w_1$, its expression is found to
be
\begin{equation}
 v_{1} \textrm{ such as }\left\{
 \begin{array}{ccc}
 v_{1}(\vec r)&=&  w_{0}(\vec r) \quad \textrm{if } r\in D_{+} \\
 v_{1}(\vec r) &=& -w_{0}(\vec r) \quad \textrm{if } r\in D_{-}
 \end{array}\right.
\end{equation}
As $w_{0}$ is orthogonal to $w_{1}$, which is the flipped mode of
$v_{0}$, one can show that $v_{0}$ is orthogonal to $v_{1}$, which
is the flipped mode of $w_{0}$ (see figure \ref{fig}). In order to
calculate the noise in the measurement using that basis, the
flipped mode is expressed as a linear combinaison of the two first
modes of the eigenbasis :
\begin{equation}
    w_{1}=\alpha v_{0}+\beta v_{1}
\end{equation}
where $\alpha=i_+-i_-$ and $\beta=2\sqrt{i_+i_-}$,
which leads to :
\begin{eqnarray}
  \langle(\delta\hat c_{1}^{\dagger}+\delta\hat
  c_{1})^{2}\rangle & = &
  \alpha^{2}\langle(\delta\hat b_{0}^{\dagger}+\delta\hat b_{0})^{2}\rangle   \nonumber \\
  & & +
  \beta^{2}\langle(\delta\hat b_{1}^{\dagger}+\delta\hat
  b_{1})^{2}\rangle  \\
  & & + 2\alpha\beta\langle(\delta\hat b_{0}^{\dagger}+\delta\hat
  b_{0})(\delta\hat b_{1}^{\dagger}+\delta\hat b_{1})\rangle
  \nonumber
\end{eqnarray}
Expressed in an eigenbasis, that do not contain the detection
mode, we see that the noise arises from the individual noise of
the two first modes and from their correlation function. In that
basis, in order to reduce the noise we have several solutions :
either the two first modes are perfectly squeezed, either they are
perfectly correlated, or any solution in between. Anyway, we can
assume that if we want to make a lot of different measurements, it
is very difficult to produce correlation between the mean field
and the different vacuum modes, hence the easiest solution is to
have the mean field squeezed, and the corresponding vacuum
squeezed. The same argument as before applies, and we show that we
still need an extra mode for each extra information.

\end{document}